\documentclass[aps,prb,preprint,superscriptaddress]{revtex4-2}

\usepackage[utf8]{inputenc}
\usepackage{textcomp} 
\usepackage{amsmath}
\usepackage{multirow}
\usepackage[export]{adjustbox}
\usepackage{hyperref}
\usepackage{xcolor}

\usepackage{booktabs}
\usepackage{siunitx}

\usepackage{graphicx}
\usepackage{amsmath}
\usepackage{bm}

\begin{document}

\title{First-principles study of doping influence on twin formation in Ni-Mn-Ga non-modulated martensite}

\author{Petr \v{S}est\'{a}k}
\email[Corresponding author: ]{psestak@it.cas.cz}
\affiliation{Institute of Thermomechanics of the Czech Academy of Sciences, Prague, Czech Republic}
\affiliation{Institute of Physical Engineering, Faculty of Mechanical Engineering, Brno University of Technology, Brno, Czech Republic}

\author{Martin Heczko}
\affiliation{Faculty of Metals Engineering and Industrial Computer Science, AGH University of Krakow, Krak\'{o}w, Poland}

\author{Ladislav Straka}
\affiliation{FZU - Institute of Physics of the Czech Academy of Sciences, Prague, Czech Republic}

\author{Alexei Sozinov}
\affiliation{Tikomat Ltd., Savonlinna, Finland}

\author{Martin Zelen\'{y}}
\affiliation{Institute of Materials Science and Engineering, Faculty of Mechanical Engineering, Brno University of Technology, Brno, Czech Republic}

\date{\today}

\begin{abstract}
We investigate how chemical substitution reshapes the energetics of twin formation in non-modulated (NM) Ni–Mn–Ga martensite. Using density functional theory, we compute generalized planar fault energy (GPFE) curves for the $(101)[10\bar{1}]$ shear system in stoichiometric Ni$_{2}$MnGa and in a set of doped supercells containing Cu, Co, Fe, or Zn on different sublattices. The GPFE landscape is used as a microscopic descriptor of twinning behavior: the first barrier reflects intrinsic stacking-fault formation (twin nucleation), whereas subsequent barriers govern twin thickening and boundary motion. We find that the impact of dopants is strongly site dependent. Substitutions Cu$\rightarrow$Mn, Cu$\rightarrow$Ni, Co$\rightarrow$Ni, and Zn$\rightarrow$Mn lower the nucleation barrier and generally soften the GPFE profile, indicating more favorable conditions for twin formation and propagation; these cases also correlate with a reduced tetragonality $c/a$, which implies a smaller twinning shear and a reduced energetic cost of twin formation. In contrast, Cu$\rightarrow$Ga, Co$\rightarrow$Mn, Co$\rightarrow$Ga, Fe$\rightarrow$Ga, and Zn$\rightarrow$Ga increase GPFE barriers and hinder twinning, even though such substitutions are often used to enhance martensite stability and raise $T_{m}$. Fe$\rightarrow$Mn leaves barrier heights largely unchanged, while Fe$\rightarrow$Ni produces an anomalous GPFE response indicative of unstable twin configurations. Finally, inspired by the nanotwinning characterization of 10M/14M modulation, we link the depth of the two-layer nanotwin minimum to modulation stability. The substitutions Fe$\rightarrow$Mn, Cu$\rightarrow$Ni, and Zn$\rightarrow$Mn result in a lower energy minimum compared to the structure without the double-layered twin. The other studied substitutions favor the twin-free NM structure.
\end{abstract}

\maketitle

\section{Introduction}
In Ni--Mn--Ga ferromagnetic shape memory alloys, large magnetocrystalline anisotropy combined with the high mobility of martensite twin boundaries enables magnetically induced reorientation (MIR) of twin variants, which results in magnetic-field-induced strains of several percent~\cite{Ullakko1996-en}. Such extraordinary properties make this material highly suitable for applications in novel actuators, sensors and energy harvesters~\cite{Acet2011-ji,Ping_Liu2009-sy,Soderberg2006-bu,Lvov2016-iy}. Reorientation-induced deformation occurs if the twinning stress is small enough to be overcome by the magnetostress~\cite{Ping_Liu2009-sy,Heczko2014-bw}. However, practical use of Ni--Mn--Ga alloys is strongly affected by both the martensitic transformation temperature and the Curie temperature. In particular, the martensitic transformation temperature of near-stoichiometric Ni$_{2}$MnGa alloys is too low, while the Curie temperature, although not particularly low, is still close to the operating temperature range and can therefore be readily exceeded under service conditions, which severely limits the functional performance of the material. It must be highlighted that these temperatures, as well as the martensite crystal structure and the ability to exhibit MIR, are highly sensitive to small changes in the chemical composition of the alloys. For example, increased Mn content shifts the martensitic transformation temperature to higher values; however, it simultaneously decreases the Curie temperature~\cite{Vasilev2004-st}. Additionally, the modulated structure of 10M martensite observed near the stoichiometric composition first transforms to modulated 14M martensite with increasing Mn content and then into non-modulated (NM) martensite for highly off-stoichiometric alloys~\cite{Heczko2014-bw}, which do not exhibit MIR due to the reported high twinning stress~\cite{Sozinov2001-lf,Okamoto2008-kb}. The modulation of the 10M and 14M structures denotes the shuffling of (110) lattice planes with different periodicity, whereas NM martensite does not exhibit shuffling, and its structure corresponds to a tetragonally distorted L2$_1$ structure ($c/a > 1$) of the parent austenite phase.

Recently published first-principles calculations of generalized-planar-fault-energy (GPFE) curves reveal that the strong increase in twinning stress and the hindered twin-boundary motion in off-stoichiometric NM martensite are associated with strong antiferromagnetic interactions between Mn-excess atoms on the Ga sublattice and other Mn atoms on their regular sublattice~\cite{Heczko2026-vx}. Because twin-boundary motion is governed by the energy barriers along the GPFE pathway, an increase in these barriers is generally expected to lead to higher twinning stress and reduced mobility, thereby suppressing MIR. 

\begin{figure}[h]
    \centering
    \includegraphics[width=1.0\linewidth]{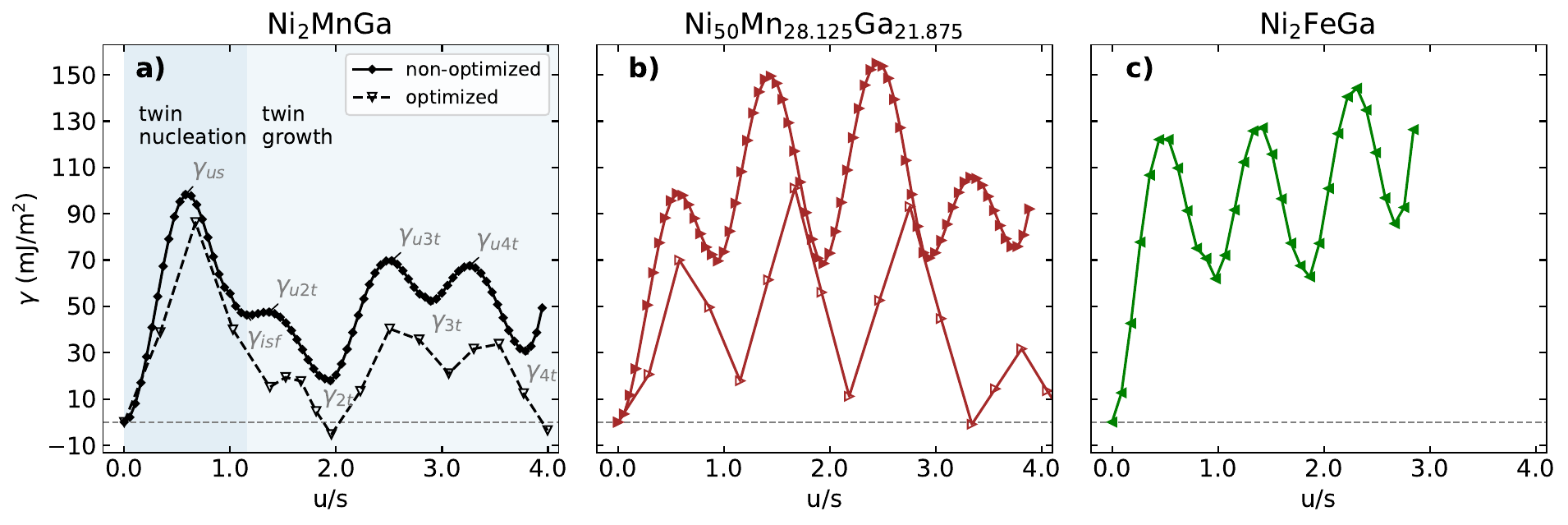}
    \caption{Generalized planar fault energy (GPFE) $\gamma$ as a function of the normalized shear displacement $u/s$ for (a) Ni$_{2}$MnGa~\cite{Heczko2024-of}, (b) Ni$_{50}$Mn$_{28.125}$Ga$_{21.875}$~\cite{Heczko2026-vx}, and (c) Ni$_{2}$FeGa~\cite{Zeleny2023-wu}. Solid lines with filled markers correspond to non-optimized atomic configurations, while dashed lines with open markers represent energies obtained after optimization. The horizontal dashed line indicates zero energy reference.}
    \label{fig:GPFE-comaprison}
\end{figure}

\begin{figure}[h]
    \centering
    \includegraphics[width=0.8\linewidth]{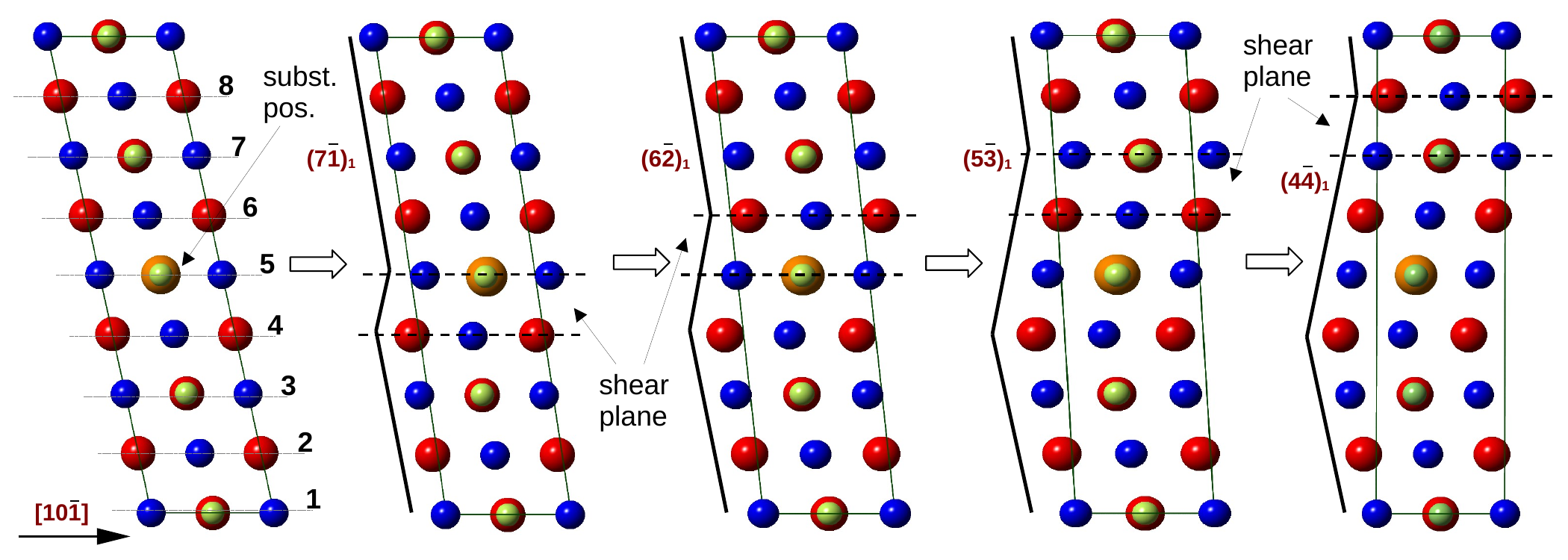}
    \caption{Schematic illustration of the nanotwin growth process, where each image represents an individual stable configuration during the nanotwin formation, i.e., a minimum on the GPFE curve. Blue, red, and green spheres represent Ni, Mn, and Ga atoms, respectively, while orange spheres represent positions of dopant atom.}
    \label{fig:twin_formation}
\end{figure}

The GPFE curves describe the energy pathways associated with twinning as a function of the normalized shear displacement $u/s$, where $u$ is the cumulative relative displacement of the sheared atomic planes and $s$ is the displacement required to complete one twinning step (see Fig.~\ref{fig:GPFE-comaprison}). The maxima on the GPFE curves represent the barriers that the crystal must overcome during shearing, whereas the minima correspond to metastable or stable stacking-fault configurations with different twin thicknesses (see Fig.~\ref{fig:twin_formation}), describing the evolution from the perfect crystal through the intrinsic stacking fault, i.e., a one-layer twin, to a multilayer twin~\cite{Wang2013-kt}. The resulting GPFE landscape can further be used as input to dislocation motion models, for example, the extended Peierls–Nabarro model for twinning stress prediction and other higher-scale deformation models~\cite{Wang2013-kt,Jagatramka2022}. Recently, the GPFE framework has also been applied as an alloy-design tool across a broader range of material systems. Combined first-principles calculations and experimental characterization have demonstrated that alloying-induced modifications of the GPFE landscape can promote deformation twinning in nanocrystalline Al alloys~\cite{Zhang2024}. Similarly, systematic first-principles studies have used GPFE analysis to screen the effects of different solute elements and their concentrations on twinning behavior in Mg alloys~\cite{Wu2024}. Furthermore, atomistic descriptions of twinning pathways based on GPFE have been employed to investigate the structure and energetics of twin interfaces in HCP metals, demonstrating the broader applicability of the framework to twin nucleation and twin-boundary migration in different HCP material systems~\cite{Gengor2023}.

For Ni--Mn--Ga alloys, the barriers along the GPFE curve increase significantly~\cite{Heczko2024-tw} if a Mn-excess atom is present near the propagating twin boundary (compare Fig.~\ref{fig:GPFE-comaprison} (a) and (b)). In contrast to other materials~\cite{Zeleny2023-wu,Ojha2014-rm} and even to other shape memory alloys~\cite{Wang2013-kt,Wang2014-ro} like for example Ni$_{2}$FeGa, the GPFE curve for stoichiometric Ni$_{2}$MnGa alloy exhibits unusual behavior (compare Fig.~\ref{fig:GPFE-comaprison} (a) and (c)), as the height of the barriers for twin propagation does not remain constant once the intrinsic stacking fault is formed, but instead oscillates depending on the twin thickness~\cite{Heczko2024-of}. The minima corresponding to even-layered twins, $\gamma _{2t}$ and $\gamma _{4t}$ exhibit significantly lower energies than odd-layered twins. If structural optimization is allowed, the energies of even-layered twins decrease even below that of the defect-free structure, indicating an exceptional stabilization of specific twin thicknesses (see dashed curve in  Fig.~\ref{fig:GPFE-comaprison} (a)). 

Importantly, the pronounced stabilization of even-layered twins inferred from the GPFE minima provides a natural microscopic basis for the lattice modulation in Ni--Mn--Ga martensites. It should be noted that stoichiometric Ni$_2$MnGa does not stabilize the NM martensite experimentally; instead, the modulated 10M structure is the stable martensitic phase at this composition. The modulation in 10M and 14M martensites may be interpreted as nanotwinning of the NM lattice along the (101) planes~\cite{Kaufmann2011-yh}, with an alternating sequence of nanotwins with thicknesses of 2 and 3 lattice planes, $(3\bar{2})_{2}$, or 2 and 5 lattice planes, $(5\bar{2})_{2}$, respectively. Thus, there is a direct relation between the low energy of the two-layered twin $\gamma _{2t}$ appearing on the GPFE curve and the stability of lattice modulation in near-stoichiometric alloys. This relation is further supported by the fact that the energy of the two-layered twin increases above the energy of the defect-free structure with increasing Mn content~\cite{Heczko2026-vx}, similarly to how the modulated structures become unstable and NM martensite is stabilized for alloys with high Mn content. Moreover, previous first-principles calculations for the stoichiometric alloy reported the lowest energy for modulated 4O martensite~\cite{Zeleny2016-en}, which consists solely of oppositely oriented two-layered twins and becomes less stable than NM martensite for off-stoichiometric alloys~\cite{Zeleny2019}. These results are also consistent with the behavior of the GPFE curves. 

A reasonable way to improve the transformation temperatures and other functional properties of the alloy is to modify its composition by doping with selected elements. It has been shown that elements such as Cu, Co, and Fe are promising candidates for tailoring the functional properties of this alloy. However, the effect of doping strongly depends on which element is replaced by the dopant in the alloy composition. Several works have been devoted to systematic investigations of the effects of dopants on $T_{m}$ and $T_{c}$~\cite{Heczko2024-tw,Li2011-fs,Kanomata2005-hj,Chen2022-sy,Kataoka2010-le,Li2011-xw,Tan2012-vf,Kopecky2021-jk,Kanomata2009-bi,Soto-Parra2010-am,Khan2005-pd,Soto2008-eh,Yu2011-dl,Liu2002,Guo2005-zm,Barton2014-ox}. The experimental results for doping on particular sublattices are summarized in Table~\ref{tab:configurations}. These observations lead to a rational strategy for the use of dopants based on the following key ideas: Co is used to replace Ni to increase $T_{c}$~\cite{Kanomata2009-bi} and decrease the $c/a$ ratio of NM martensite~\cite{Soroka2018-ci}, whereas Cu at the expense of Mn and Ga should stabilize the martensitic phase and increase $T_{m}$~\cite{Li2011-xw,Kataoka2010-le}. Careful balancing of dopant concentrations allows preparation of the Ni$_{46}$Co$_{4}$Mn$_{24}$Ga$_{22}$Cu$_{4}$ alloy with an NM structure stable at room temperature, which exhibits sufficiently high $T_{m}$ and $T_{c}$. It also exhibits a remarkably high magnetic-field-induced strain of 12\% due to a sufficiently low twinning stress that enables MIR~\cite{Sozinov2013-si}. The reported twinning stress is significantly lower than the twinning stress reported for NM martensite stabilized by excess Mn and is attributed to two factors: (i) a reduced tetragonal distortion of the NM structure ($c/a$ $\approx$ 1.15 in the Co--Cu-doped alloy versus $\approx$ 1.21 in the Mn-excess variant)~\cite{Soroka2018-ci} and (ii) a reorientation of the softest elastic shearing modes toward the $\{101\}\langle101\rangle$ directions, which is the shear system for twin formation in NM martensite~\cite{Bodnarova2020-bg}. However, the presence of Cu in the alloy reduces magnetocrystalline anisotropy~\cite{Zeleny2020-ll}, which decreases the actuation stress provided by the alloy.

On the other hand, substituting Fe for Ga leads to very low twinning stress in modulated martensite; in particular, the lowest twinning stress was observed for 14M martensite in an alloy doped with 5\% Fe~\cite{Sozinov2020-vl}. In addition, Fe substitution is responsible for extending the thermodynamic stability range of modulated structures~\cite{RIOLOPEZ2026186643}, although improper balancing of dopants between the Ga and Mn sublattices can even decrease $T_{m}$. Alloys simultaneously doped with Cu, Co, and Fe also exhibit promising properties, such as increased $T_{m}$ and $T_{c}$~\cite{Perez-Checa2019-ad,Perez-Checa2017-hz,Perez-Checa2018-kr}. Another promising dopant is Zn, because substituting Zn for Ga also increases $T_{m}$~\cite{Barton2014-ox}, while reducing magnetocrystalline anisotropy less strongly than Cu does~\cite{Janovec2020-ko}. First-principles simulations predict that Zn on the Ga sublattice could increase $T_{c}$, but Zn on the Mn sublattice decreases both $T_{m}$ and $T_{c}$~\cite{Janovec2020-ko}. However, fabrication, and consequently experimental study, of Ni--Mn--Ga--Zn systems with higher Zn content is strongly limited by the intensive evaporation of Zn during alloying~\cite{Barton2014-ox}, although this limitation might potentially be mitigated by alternative fabrication routes that avoid conventional melting.

The purpose of this study is to assess the influence of the dopant elements Cu, Co, Fe, and Zn on the profile of the GPFE curve of NM martensite. To investigate this influence, we employ an atomistic model introduced in our previous works~\cite{Zeleny2023-wu,Heczko2024-of}. This model allows us to estimate the ability of doped alloys to nucleate twins and support their subsequent growth, since higher barriers along the curve hinder twin-boundary motion and increase twinning stress, whereas reduced barriers should correspond to lower twinning stress and support MIR. Moreover, increased or decreased stability of particular twin configurations, i.e., more or less pronounced minima along the curve, may indicate the stability of lattice modulation in the martensitic phase. For doping by Cu, Co, and Fe, we take into account all possible substitutional sites, i.e., replacement of Ni, Mn, or Ga atoms, because a large number of experimental works are available for comparison. For doping by Zn, we consider only replacement of Mn and Ga. Thus, in total we evaluate GPFE curves for eleven alloy compositions, which significantly contributes to understanding how doping affects the properties of Ni--Mn--Ga magnetic shape memory alloys.

\begin{table}[h]
    \centering
    \caption{Overview of the considered doping configurations, together with experimentally determined effects on $T_{m}$ and $T_{c}$ ($\uparrow\uparrow$ -- strong increase, $\uparrow$ -- increase, $\nearrow$ -- small increase, $\searrow$ -- small decrease, $\downarrow$ -- decrease, $\downarrow\downarrow$ -- strong decrease) and site preference according to~\cite{Li2011} and~\cite{Janovec2020-ko}.}
    \label{tab:configurations}
    \begin{tabular}{ccccc}
        \hline
       substitution & formula & $T_{m}$ & $T_{c}$ & site preference \\
        \hline
        Cu$\rightarrow$Mn$^{a,b,c}$ & Ni$_{50}$Mn$_{21.875}$Cu$_{3.125}$Ga$_{25}$ & $\uparrow$ & $\downarrow\downarrow$ & direct \\
       Cu$\rightarrow$Ga$^{a,d,e}$ & Ni$_{50}$Mn$_{25}$Ga$_{21.875}$Cu$_{3.125}$ & $\uparrow\uparrow$ & $\downarrow$ & direct \\
       Cu$\rightarrow$Ni$^{a,f,g}$ & Ni$_{46.875}$Cu$_{3.125}$Mn$_{25}$Ga$_{25}$ & $\downarrow\downarrow$ & $\uparrow$ & direct \\
        \hline
       Co$\rightarrow$Mn$^{h,i}$ & (Ni$_{46.875}$Co$_{3.125}$)(Mn$_{21.875}$Ni$_{3.125}$)Ga$_{25}$ & $\uparrow$ & $\nearrow$ & antisite \\
       Co$\rightarrow$Ga$^{g,h}$ & (Ni$_{46.875}$Co$_{3.125}$)Mn$_{25}$(Ga$_{21.875}$Ni$_{3.125}$) & $\uparrow\uparrow$ & $\downarrow$ & antisite \\
       Co$\rightarrow$Ni$^{h,j,k}$ & Ni$_{46.875}$Co$_{3.125}$Mn$_{25}$Ga$_{25}$ & $\downarrow\downarrow$ & $\uparrow$ & direct \\
        \hline
       Fe$\rightarrow$Mn$^{h,l,m}$ & Ni$_{50}$Mn$_{21.875}$Fe$_{3.125}$Ga$_{25}$ & $\searrow$ & $\nearrow$ & direct \\
       Fe$\rightarrow$Ga$^{h,l,n}$ & Ni$_{50}$(Mn$_{21.875}$Fe$_{3.125}$)(Ga$_{21.875}$Mn$_{3.125}$) & $\uparrow$  & $\nearrow$ & antisite \\
       Fe$\rightarrow$Ni$^{h,k,m,o}$ & Ni$_{46.875}$Fe$_{3.125}$Mn$_{25}$Ga$_{25}$ & $\downarrow$ & $\uparrow$ & direct \\
        \hline
       Zn$\rightarrow$Mn & Ni$_{50}$Mn$_{21.875}$Zn$_{3.125}$Ga$_{25}$ & $\downarrow$* & $\downarrow\downarrow$* & direct \\
       Zn$\rightarrow$Ga$^{p}$ & Ni$_{50}$Mn$_{25}$Ga$_{21.875}$Zn$_{3.125}$ & $\uparrow$ & $\searrow$* & direct \\
        \hline
       \multicolumn{5}{l}{ Experimental results for $T_{m}$ and $T_{c}$ from $^{a}$\cite{Heczko2024-tw}, $^{b}$\cite{Chen2022-sy}, $^{c}$\cite{Kataoka2010-le}, $^{d}$\cite{Li2011-xw}, $^{e}$\cite{Tan2012-vf}, $^{f}$\cite{Li2011-fs}, $^{g}$\cite{Kanomata2005-hj}, $^{h}$\cite{Kopecky2021-jk}, $^{i}$\cite{Khan2005-pd},} \\ 
       \multicolumn{5}{l}{$^{j}$\cite{Kanomata2009-bi}, $^{k}$\cite{Soto-Parra2010-am}, $^{l}$\cite{Soto2008-eh}, $^{m}$\cite{Liu2002}, $^{n}$\cite{Guo2005-zm}, $^{o}$\cite{Yu2011-dl}, $^{p}$\cite{Barton2014-ox}, * denotes results of first-principles calculations~\cite{Janovec2020-ko}.}  \\ 
    \end{tabular}
    \label{tab:doping}
\end{table}

\section{Methods}

\begin{figure}[h]
    \centering
    \includegraphics[width=0.5\linewidth]{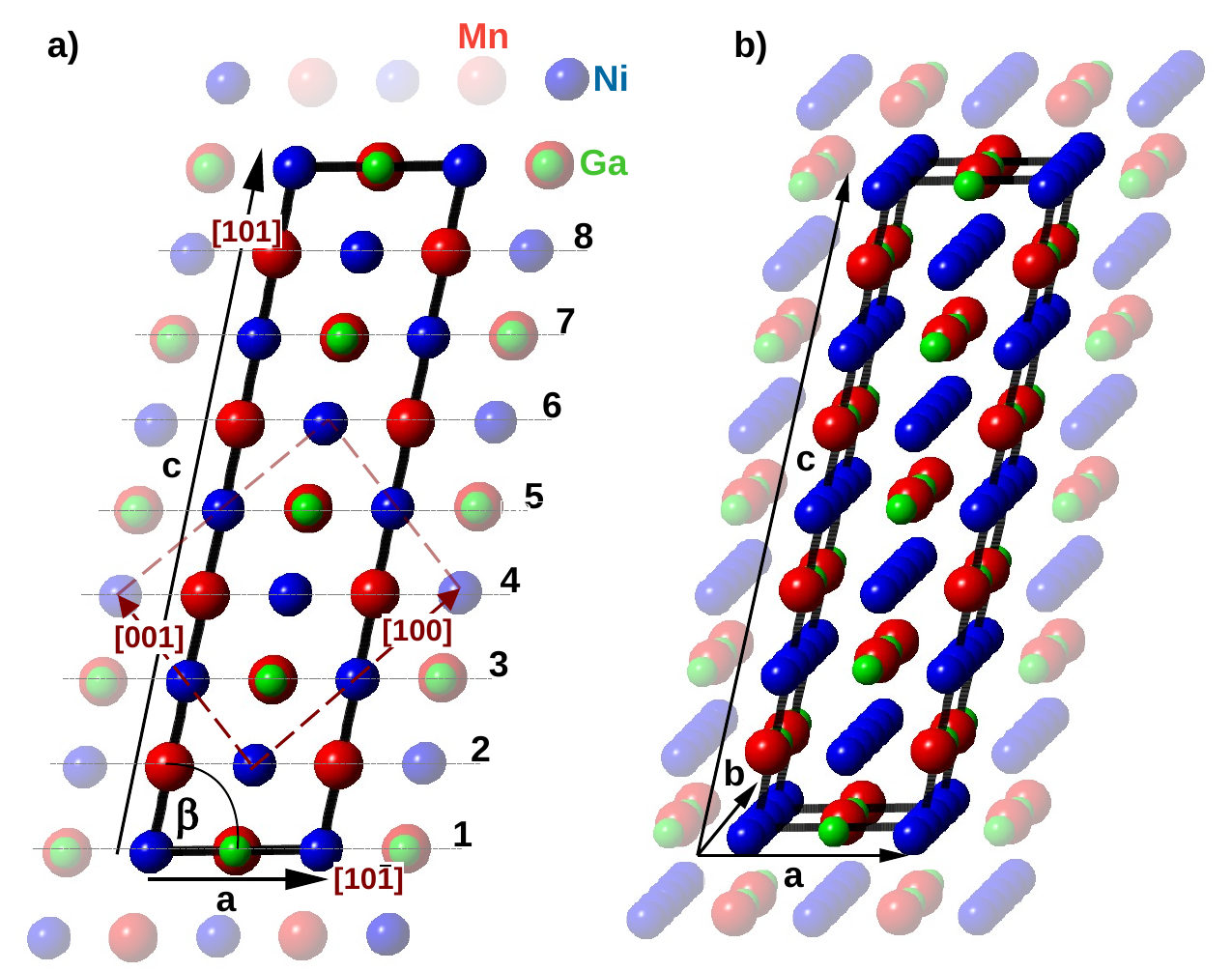}
    \caption{Simulation cell used in the present atomistic model for stoichiometric Ni$_{2}$MnGa NM martensite.  (a) View of the simulation cell from [010] direction. (b) Three-dimensional view of the simulation cell highlighting the periodic boundary conditions. Blue, red, and green spheres represent Ni, Mn, and Ga atoms, respectively. The translation vectors are indicated by arrows, and the simulation cell used in the present atomistic model is marked by a thick black line while the dashed lines represent the conventional tetragonally distorted L2$_{1}$ cell of NM martensite.}
    \label{fig:gpfe_scheme}
\end{figure}

The obtained results were determined from first principles calculations employing density functional theory (DFT) and its particular implementation in the VASP code \cite{Kresse1993,Kresse1996}. We used the projector augmented-wave method \cite{Kresse1999} and the exchange-correlation energy was evaluated within the generalized gradient approximation parametrized by Perdew, Burke, and Ernzerhof (PBE) \cite{Perdew1996}. The magnetic ordering was considered using spin-polarized calculations. The plane-wave cut-off energy was set to 500~eV and the first Brillouin zone was sampled using a $13 \times 4 \times 10$ $\Gamma$-centered Monkhorst--Pack mesh that corresponded to the DFT supercell with 32 atoms consisting of eight (101) planes of NM martensite, as illustrated in Fig.~\ref{fig:gpfe_scheme}. Calculations were performed using the Methfessel--Paxton smearing scheme with a smearing width of 0.2 eV. The energy was converged in the self-consistent cycle to reduce its fluctuations below $10^{-7}$~eV. The doped initial structures were fully relaxed prior to the shearing procedure until the residual forces on all atoms were below 1 meV/\AA~and the stress tensor components were reduced below 50 MPa. To compute the GPFE curves of doped Ni--Mn--Ga NM martensite, a supercell geometry was constructed to allow deformation along the $(101)[10\bar{1}]$ shear system. A monoclinic supercell was defined with crystallographic axes aligned along $[10\bar{1}]$, $[101]$, and $[010]$ directions as illustrated in Fig.~\ref{fig:gpfe_scheme}. The full elastic tensor was calculated after structural relaxation using the stress--strain method described in Refs.~\cite{Yu2010-aj,Zhou2013-uq}. For each applied external strain of $2\%$, all atomic positions were fully relaxed. The shear modulus corresponding to the investigated shear system, $G_{(101)[10\bar{1}]}$, was subsequently obtained using the in-house \textsc{ElaStr} tool~\cite{Zeleny_undated-wc}.

The deformation pathway associated with nanotwin formation was described using the simple shear model, which has already been introduced and compared with two alternative approaches in our previous work \cite{Heczko2024-of}. In this framework, the shear deformation is implemented as a cumulative displacement applied to selected (101) lattice planes along the $[10\overline{1}]$ direction, while the lattice volume is maintained constant throughout the entire process and there are no optimizations of the lattice parameters or ionic positions during the shear. The displacement applied to the upper four lattice planes was defined as

\begin{equation}
u = \frac{|\mathbf{s}|}{n_{\mathrm{GPFE}}},
\end{equation}

where $\mathbf{s}$ denotes the shearing vector oriented along the $[10\bar{1}]$ direction, and $n_{\mathrm{GPFE}}$ represents the number of incremental deformation steps between successive minima of the GPFE curve corresponding to metastable twin configurations. The shear vector $\mathbf{s}$ describes the relative displacement between adjacent lattice planes required to generate a perfectly mirror-symmetric twin configuration in the tetragonal lattice \cite{Zeleny2023-wu,Heczko2024-of}. The ideal magnitude of the shear vector $|\mathbf{s}|$ in the body-centered tetragonal (bct-like) lattice of NM martensite can be expressed as

\begin{equation}
|\mathbf{s}| = \frac{1}{2}\left(\frac{(c/a)^2 - 1}{(c/a)^2 + 1}\right)\cdot |[10\bar{1}]|,
\end{equation}

where $c/a$ denotes the tetragonal ratio of the lattice. The expression follows from the crystallographic description of twinning in tetragonal systems \cite{Zeleny2023-wu}.

This applied displacement produces the first energy maximum on the GPFE curve $\gamma _{us}$, followed by a local minimum $\gamma _{isf}$ associated with a metastable stacking-fault configuration which can be denoted for our supercell as $(7\bar{1})_{1}$ in nanotwinning notation. This minimum is then adopted as the initial state for the subsequent shear increment, in which the adjacent atomic plane is displaced, again giving rise to a new maximum $\gamma _{u2t}$ and a subsequent minimum $\gamma _{2t}$ on the GPFE profile. By iteratively applying this procedure to successive neighboring planes, the second $(6\bar{2})_{1}$, third $(5\bar{3})_{1}$, and finally the fourth nanotwin layers are formed, resulting in the fully developed four-layer nanotwin structure that is displayed in  Fig.~\ref{fig:twin_formation} and marked as $(4\bar{4})_{1}$. The energy profile obtained directly from this pathway, without any subsequent structural relaxation, is hereafter referred to as the non-optimized GPFE and is depicted in all figures as a solid line.

To obtain a more realistic description of nanotwin formation, the local minima $\gamma _{isf}$, $\gamma _{2t}$, $\gamma _{3t}$, $\gamma _{4t}$ previously identified on the non-optimized GPFE surface were further refined via additional structural relaxations using the GADGET tool~\cite{Bucko2005}, while maintaining the same simple shear constraints, i.e. the cell size in shear plane and height of the cell perpendicular to shear plane was still kept constant. As discussed in our previous work \cite{Heczko2024-of}, the direct displacement pathway yields only an approximate representation of the underlying energy landscape, because local structural relaxations in the vicinity of the twin boundary can slightly modify both the position and the depth of the minima. The constrained refinement thus enables an accurate determination of the true equilibrium configurations associated with individual nanotwin layers.

In contrast, direct relaxation of configurations near the energy maxima is not suitable, as such states tend to relax towards neighboring minima. For this reason, the corresponding transition states were obtained using the generalized solid-state nudged elastic band (G-SSNEB) method~\cite{Sheppard2012, HjorthLarsen2017}, which enables reliable determination of the minimum-energy path and associated energy barriers. The resulting energy profile is denoted in this work as the optimized GPFE and is shown by dashed lines. For the present simulations, the minimum-energy paths were determined using five images, including the initial and final configurations, with a spring constant of $k=5.0$~eV/\AA$^{2}$. The paths were optimized using the climbing-image method until the maximum force fell below 0.05~eV/\AA.  

To evaluate how chemical substitution modifies the GPFE landscape, we introduced a single dopant atom into the simulation cell, so that the concentration of dopant element was kept constant at 3.125 at.\% in all cases. Owing to the periodic boundary conditions, the resulting model describes a periodically repeated dopant arrangement rather than a statistically random solid solution. An important aspect that must be considered in our configurations is the site preference of the dopant atoms. This issue for 3d dopants has been extensively analyzed by Li et al.~\cite{Li2011}. For instance, substituting a Ga atom with Fe can give rise to multiple atomic configurations. The most straightforward case is the direct substitution of Ga by Fe. However, an alternative configuration is also possible, in which the Ga site is occupied by a neighboring Mn atom, while the vacated Mn site is subsequently occupied by the Fe dopant. Such antisite as well as direct configurations were analyzed in detail for Co, Cu, and Fe in~\cite{Li2011} and for Zn in~\cite{Janovec2020-ko}. In the present study, we adopted only the configurations from the aforementioned works that correspond to the lowest-energy state. These configurations therefore provide controlled local atomic environments suitable for comparing the effects of different dopants and substitution mechanisms. Specifically, antisite occupancy was reported for Co replacing Mn and Ga (Co$\rightarrow$Mn, Co$\rightarrow$Ga), where Co prefers to occupy the Ni sublattice, and for Fe replacing Ga (Fe$\rightarrow$Ga), where Fe prefers to occupy the Mn sublattice, as in the example described above. Moreover, the strong site preferences of Co and Fe leading to antisite arrangements have also been confirmed experimentally~\cite{RIOLOPEZ2026186643, Ayila2012-ps}. The other doping combinations studied in this work prefer direct substitution. Site preference for all configurations, along with the notation used to label them, are summarized in Table~\ref{tab:configurations}.

In the 32-atom supercell, the dopant atom always substitutes the host atom in the fifth plane (see Fig.~\ref{fig:twin_formation}). When antisite occupancy is energetically more favorable, we constructed the corresponding antisite configuration such that the required accompanying swap is also located within the fifth plane, to keep the dopant environment comparable across cases. For example, in the aforementioned antisite configuration for Fe doping at the expense of Ga, the fifth plane contains two Ni atoms, an Fe atom on an Mn site (Fe$_{Mn}$), and an Mn atom on a Ga site (Mn$_{Ga}$). The Mn$_{Ga}$ atom exhibits a magnetic moment opposite to Mn atoms on their regular sublattice due to strong antiferromagnetic coupling. In contrast, when Ni atoms occupy Mn or Ga sublattices (Ni$_{Mn}$ or Ni$_{Ga}$), they do not change the orientation of their magnetic moment. For compositions in which Ni atoms are directly substituted, we found that the structural stability is very sensitive to an unrealistically high local concentration of dopant within a single layer. In these cases, we doubled the supercell in the x-direction to obtain a 64-atom cell. Only one of four Ni atoms was replaced in the fifth layer, and a second Ni atom was replaced in the second layer, while keeping the overall dopant concentration at 3.125 at.\%. This configuration reduces the artificial in-plane clustering of dopants present in the 32-atom cell, but it still represents a periodically ordered distribution rather than a statistically random arrangement. The separation of the dopants between the second and fifth layers provides a more dilute local environment while preserving the same overall composition.

\section{Results}

\begin{figure}[ht!]
  \centering
  \includegraphics[width=\textwidth]{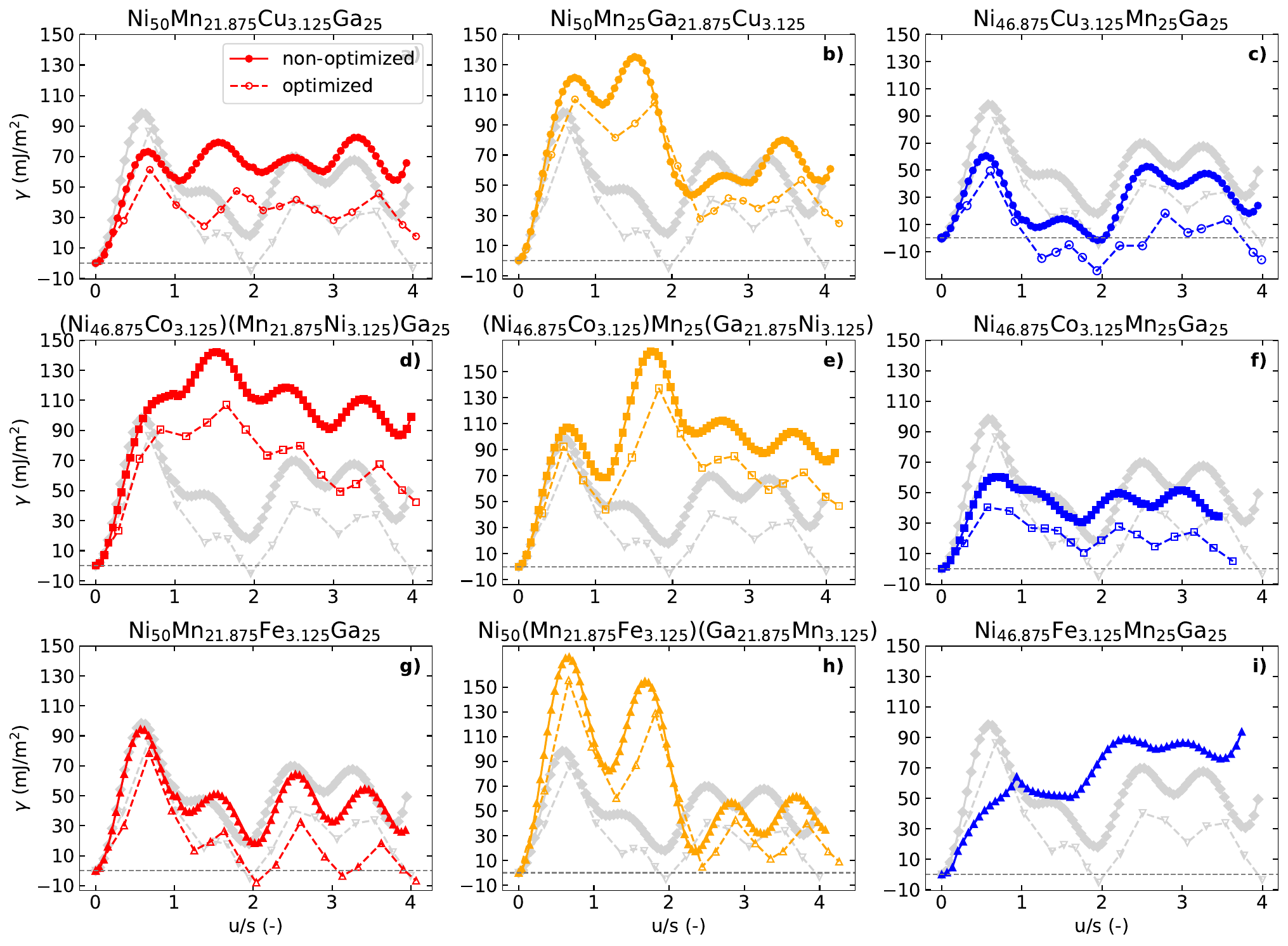}
  \caption{The generalized planar fault energy (GPFE) curves obtained for (a)--(c) Cu-, (d)--(f) Co-, and (g)--(i) Fe-doped systems. The columns correspond to substitution of the host species Mn in panels (a), (d), and (g), Ga in panels (b), (e), and (h), and Ni in panels (c), (f), and (i) of Ni$_2$MnGa. Solid curves correspond to non-optimized configurations, whereas dashed curves represent the optimized GPFE profiles obtained using the GADGET algorithm together with the nudged elastic band (NEB) method. For comparison, the non-optimized and optimized GPFE curves of stoichiometric Ni$_2$MnGa are displayed in gray in each panel. Individual dopant configurations are further distinguished by color, as indicated in the respective panels.}
  \label{fig:main}
\end{figure}

To reveal how chemical substitution affects nanotwin formation in NM martensite, we calculate GPFE curves for Cu-, Co-, Fe-, and Zn-doped Ni$_{2}$MnGa across all selected substitutional configurations. Fig.~\ref{fig:main} presents the results for Cu, Co, and Fe dopants. For each dopant, both the non-optimized and optimized GPFE profiles (solid and dashed lines, respectively) are evaluated and compared with the profiles for stoichiometric Ni$_{2}$MnGa alloy~\cite{Heczko2024-of} from Fig.~\ref{fig:GPFE-comaprison} (a), which are shown in the background in gray. Because these profiles serve as a reference for subsequent comparisons, we first briefly analyze their behavior along the shear path and the effect of optimization. As discussed in our previous studies~\cite{Heczko2026-vx}, the most significant features of the curve are pronounced minima corresponding to two-layer and four-layer nanotwins, $\gamma _{2t}$ and $\gamma _{4t}$. In the optimized curves, these minima become even more energetically favorable than the initial defect-free structure. The formation of the intrinsic stacking fault (i.e., a one-layer nanotwin) is the most energetically demanding step, as it is associated with the highest energy maximum $\gamma _{us}$ in both the non-optimized and optimized GPFE curves. After formation of the intrinsic stacking fault in stoichiometric Ni$_{2}$MnGa, a local minimum $\gamma _{isf}$ appears on the GPFE curve. However, this minimum is followed by only a small barrier ($\gamma _{u2t}-\gamma _{isf}$), indicating that formation of the first two nanotwin layers may effectively proceed within a single shearing process. A similar behavior is observed for subsequent layers: while formation of the three-layer nanotwin presents a more pronounced energy barrier ($\gamma _{u3t}-\gamma _{2t}$), the barrier associated with formation of the four-layer nanotwin ($\gamma _{u4t}-\gamma _{3t}$) is significantly lower. It is evident that the additional structural optimization performed using the aforementioned procedure substantially lowers the GPFE energy profile and should not be omitted for a reliable evaluation of dopant effects; therefore, we employ both approaches in the assessment of doped systems. On the other hand, the effect of optimization appears to be mainly quantitative rather than qualitative, since the relative comparison of non-optimized and optimized curves between stoichiometric and doped alloys leads to the same qualitative prediction for twinning behavior. Therefore, in the following we discuss the general trends without distinguishing between the non-optimized and optimized curves. The numerical values of each maximum along the non-optimized curves, and the corresponding optimized values (in brackets), are listed in Table~\ref{tab:gpfe_peaks}.

\begin{table}[h]
\centering
\caption{Local maxima on the GPFE curves (see Fig.~\ref{fig:GPFE-comaprison} (a) for their definition), tetragonal $c/a$ ratio, volume per atom $V_{at}$ and the predicted effect on twinning for the considered alloys. Values obtained after optimization are given in parentheses. N/A indicates that the corresponding maximum is absent from the GPFE curve.}
\label{tab:gpfe_peaks}
\begin{tabular}{l@{\hspace{6pt}}l @{\hspace{6pt}}l @{\hspace{6pt}}l @{\hspace{6pt}}l @{\hspace{10pt}}r @{\hspace{10pt}}c l}
\hline
\multirow{2}{*}{substitution} & \multicolumn{4}{c}{GPFE maxima (mJ/m$^{2}$)} & \multirow{2}{*}{$c/a$} & \multirow{2}{*}{$V_{at}$ (\AA$^{3}$)} & \multirow{2}{*}{twinning} \\
\cline{2-5}
 & $\gamma_{us}$ & $\gamma_{u2t}$ & $\gamma_{u3t}$ & $\gamma_{u4t}$ &  &  &  \\
\hline

Ni$_{2}$MnGa & 98.3 (86.3) & 47.5 (19.4) & 69.7 (40.3) & 67.6 (33.7) & 1.252 & 12.24 & -- \\
\hline

Cu$\rightarrow$Mn & 73.1 (61.3) & 79.2 (47.3) & 69.3 (41.7) & 82.4 (45.6) & 1.239 & 12.14 & support \\
Cu$\rightarrow$Ga & 121.4 (107.0) & 135.2 (104.9) & 56.4 (41.5) & 80.0 (53.5) & 1.254 & 12.11 & hinder \\
Cu$\rightarrow$Ni & 60.6 (49.5) & 14.2 (-5.0) & 52.7 (18.2) & 47.5 (13.4) & 1.238 & 12.24 & support \\
\hline

Co$\rightarrow$Mn & 114.4 (90.5) & 142.3 (107.1) & 118.5 (79.9) & 110.8 (67.6) & 1.243 & 12.08 & hinder \\
Co$\rightarrow$Ga & 106.8 (92.4) & 165.6 (137.1) & 112.3 (85.0) & 103.4 (72.6) & 1.256 & 12.05 & hinder \\
Co$\rightarrow$Ni & 60.4 (40.4) & N/A (N/A) & 50.0 (27.6) & 51.7 (24.3) & 1.217 & 12.17 & support \\
\hline

Fe$\rightarrow$Mn & 94.4 (78.5) & 51.1 (26.4) & 64.4 (32.5) & 54.6 (18.3) & 1.256 & 12.14 & negligible effect \\
Fe$\rightarrow$Ga & 174.2 (155.5) & 154.6 (128.7) & 57.0 (42.6) & 61.7 (39.8) & 1.279 & 12.09 & hinder \\
Fe$\rightarrow$Ni & 64.2 & N/A & 89.1 & 86.5 & 1.195 & 12.16 & -- \\
\hline

Zn$\rightarrow$Mn & 48.1 (35.0) & N/A (N/A) & 42.8 (3.2) & 58.1 (16.3) & 1.227 & 12.20 & support \\
Zn$\rightarrow$Ga & 110.0 (97.5) & 80.4 (N/A) & 55.0 (34.0) & 75.3 (44.2) & 1.247 & 12.16 & hinder \\
\hline
\end{tabular}
\end{table}

\subsection{Cu doping}

Cu$\rightarrow$Mn substitution (Fig.~\ref{fig:main} (a)) leads to a pronounced reduction of the first energy barrier $\gamma _{us}$ together with shallower and nearly equivalent local minima along the GPFE curve, suggesting facilitated nucleation of the initial nanotwin layers. The subsequent barrier heights also become more comparable in magnitude than in stoichiometric Ni$_2$MnGa, indicating an overall softening of the energetics associated with nanotwin propagation and resulting in a considerably smoother GPFE profile. Consequently, the GPFE profile becomes much more similar to those of common metals and other alloys~\cite{Zeleny2023-wu,Ojha2014-rm,Wang2013-kt,Wang2014-ro} (compare with Fig.~\ref{fig:GPFE-comaprison} (c)). In contrast, Cu$\rightarrow$Ga substitution (Fig.~\ref{fig:main} (b)) increases the GPFE maxima, especially $\gamma _{us}$ and $\gamma _{u2t}$, which are associated with formation of the intrinsic stacking fault and the two-layer nanotwin, respectively. This indicates a less favorable energy landscape for nanotwin nucleation. Cu$\rightarrow$Ni substitution (Fig.~\ref{fig:main} (c)) produces a significant reduction of GPFE barriers across multiple nanotwin configurations, accompanied by a smoother energy profile, indicating favorable conditions for nanotwin growth. Moreover, the second minimum exhibits lower energy than the initial defect-free structure even for the non-optimized curve. This behavior becomes even more pronounced when structural optimization is employed: not only the two-layer twin is significantly stabilized, but the four-layer twin and the intrinsic stacking fault also exhibit energies lower than the initial defect-free structure. Furthermore, even the maximum between the intrinsic stacking fault and the two-layer twin, $\gamma_{u2t}$, lies energetically below the initial NM martensite structure. This unusual feature is related to the strong stabilization of nanotwinned configurations and is discussed in more detail in Sec.~\ref{sec:discussion}. On the other hand, a non-negligible barrier, given by $(\gamma_{u2t}-\gamma_{isf})$, still exists between these two configurations. 

\subsection{Co doping}

As Co exhibits a strong preference to occupy the Ni sublattice~\cite{Li2011}, the Co$\rightarrow$Mn and Co$\rightarrow$Ga cases were therefore evaluated in the antisite configuration. Co$\rightarrow$Mn doping (Fig.~\ref{fig:main} (d)) generally shifts the GPFE energy profile up, although the barrier between the two- and three-layer twins ($\gamma _{u3t}-\gamma _{2t}$) appears smaller than in the stoichiometric reference. However, the energies of the second and subsequent GPFE maxima are significantly increased, leading to greater resistance to nanotwin growth. Substitution of Co$\rightarrow$Ga (Fig.~\ref{fig:main} (e)) results in the largest GPFE barrier for formation of the two-layer nanotwin ($\gamma _{u2t}-\gamma _{isf}$). When Co substitutes Ni (Co$\rightarrow$Ni, Fig.~\ref{fig:main} (f)), the GPFE curve shows a substantially lower energy barrier $\gamma_{us}$ for twin nucleation and an overall lowering of the profile for formation of the remaining layers. Moreover, the local minimum corresponding to the intrinsic stacking fault, $\gamma_{isf}$, disappears, suggesting that formation of the first two nanotwin layers may proceed without a distinct metastable intermediate state. These features indicate an overall softening effect induced by Co substitution on the Ni sublattice

\subsection{Fe doping}

The GPFE profile for Fe→Mn substitution (Fig.~\ref{fig:main} (g)) remains close to the stoichiometric Ni$_{2}$MnGa GPFE curve for formation of the first nanotwin layer and modifies only slightly the energetics of the later minima and maxima. In particular, in the optimized GPFE curve the minima for two- and three-layer twins, $\gamma _{2t}$ and $\gamma _{3t}$, become more pronounced, which indicates higher stability of these configurations compared to the stoichiometric alloy. In general, despite this local stabilization, the influence of Fe$\rightarrow$Mn substitution can be considered almost negligible. Fe$\rightarrow$Ga substitution (Fig.~\ref{fig:main} (h)), which exhibits a strong preference for antisite defects with opposite orientation of Mn$_{Ga}$ magnetic moments, leads to a substantial modification of the GPFE curve, including increased energy barriers and a reshaping of the energy landscape for higher-order nanotwin layers. The role of this magnetic arrangement was investigated explicitly in our recent study of Mn-excess Ni–Mn–Ga martensite \cite{Heczko2026-vx}. In that work, GPFE curves calculated for different magnetic configurations showed that antiparallel alignment of excess Mn$_{Ga}$ moments, corresponding to the magnetic ground state, increases the barriers for twin nucleation and propagation as shown in Fig.~\ref{fig:GPFE-comaprison} (b), whereas the energetically less favorable parallel alignment lowers them. The increase was found to depend strongly on the local position of Mn$_{Ga}$ relative to the propagating twin. These findings suggest that the antiparallel ordering of Mn$_{Ga}$ moments may also contribute to the high barriers obtained for Fe$\rightarrow$Ga substitution. However, because the present study does not compare alternative magnetic configurations for Fe$\rightarrow$Ga, its magnetic contribution cannot be separated from the simultaneous changes in the local chemical environment. The high value of $\gamma_{us}$ therefore cannot be attributed solely to the antiparallel magnetic ordering. The pronounced modification of the GPFE curve is also localized around the dopant-affected region: once the moving twin-boundary plane has passed through the plane containing the dopant-related atomic rearrangement, the barriers for subsequent twin propagation become very similar to those of the stoichiometric case. In contrast, Fe$\rightarrow$Ni substitution (Fig.~\ref{fig:main} (i)) results in a comparatively smooth GPFE profile with moderately reduced barriers for later nanotwin layers, suggesting a less restrictive effect on nanotwin propagation compared to Mn and Ga substitution. However, we note that the overall shape of the GPFE curve exhibits several features that do not appear in other doping cases, namely a very sharp first maximum or even a discontinuity on the GPFE curve. Furthermore, the first and third minima are barely recognizable. This indicates very low stability of these configurations and does not allow us to calculate the optimized GPFE curve. This behavior is consistent with the fact that experimentally prepared alloys with Fe substituted for Ni do not exhibit a martensitic transformation~\cite{Kopecky2021-jk}.

\subsection{Zn doping}

\begin{figure}[ht!]
  \centering
  \includegraphics[width=0.85\linewidth]{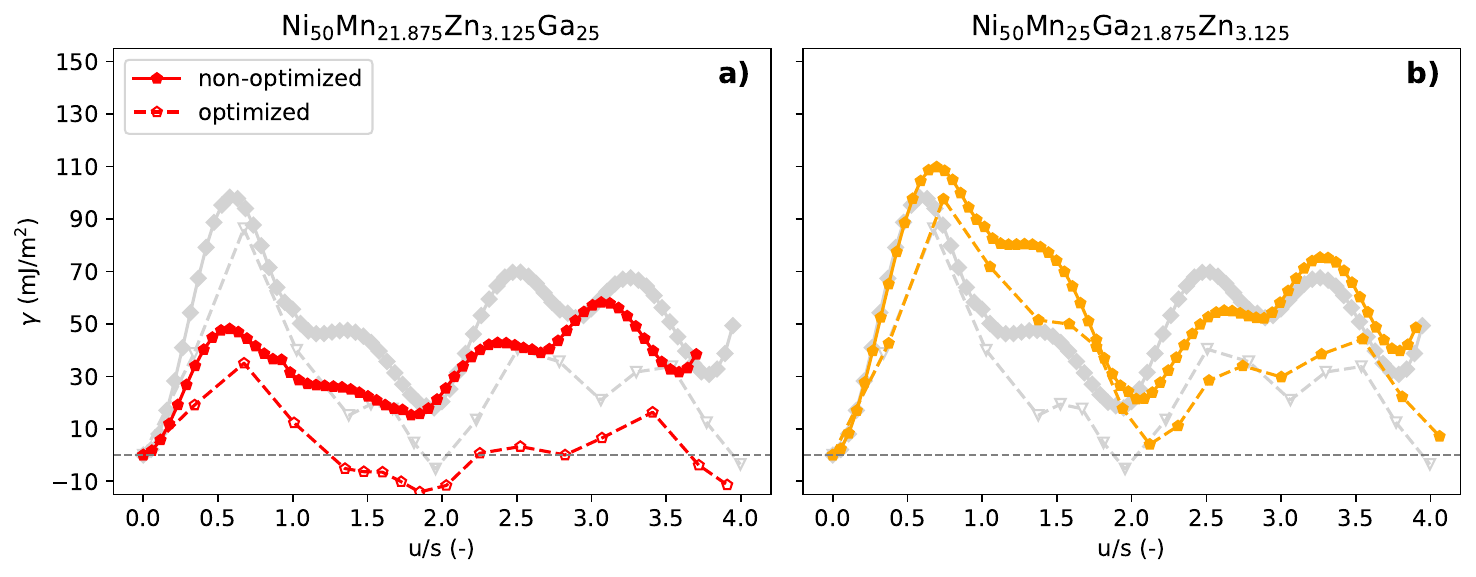}
  \caption{The generalized planar fault energy (GPFE) curves obtained for Zn-doped Ni--Mn--Ga NM martensite with (a) Zn$\rightarrow$Mn and (b) Zn$\rightarrow$Ga substitution. Solid curves correspond to non-optimized configurations, whereas dashed curves represent the optimized GPFE profiles obtained using the GADGET algorithm together with the nudged elastic band (NEB) method. For comparison, the non-optimized and optimized GPFE curves of stoichiometric Ni$_2$MnGa are displayed in gray in each panel.}
  \label{fig:Zn}
\end{figure}

Fig.~\ref{fig:Zn} shows the GPFE profiles for Mn and Ga substitution by Zn, together with the stoichiometric Ni$_{2}$MnGa profiles for comparison. For Zn$\rightarrow$Mn substitution (Fig.~\ref{fig:Zn} (a)), a pronounced reduction of the first GPFE barrier $\gamma_{us}$ is observed compared to the stoichiometric alloy. Moreover, the local minimum corresponding to the intrinsic stacking fault, $\gamma_{isf}$, is absent already in the non-optimized GPFE curve and remains absent after optimization, indicating that this configuration is not metastable. At the same time, the minimum corresponding to the two-layer twin, $\gamma_{2t}$, is significantly deepened after optimization. This indicates strong stabilization of the early stages of nanotwin formation, despite the absence of a metastable intrinsic stacking-fault configuration. In contrast, Zn substitution at the Ga site (Zn$\rightarrow$Ga, Fig.~\ref{fig:Zn} (b)) leads to an overall increase of the GPFE barriers, particularly $\gamma_{us}$ for twin nucleation, and preserves a steeper energy landscape even after optimization. The $\gamma_{isf}$ minimum is also absent for Zn$\rightarrow$Ga, indicating that the intrinsic stacking-fault configuration is not metastable for either Zn substitution. These results demonstrate that the effect of Zn doping on nanotwin energetics is strongly site-dependent, with Mn substitution facilitating twin nucleation and early growth, while Ga substitution has the opposite effect.

\section{Discussion}
\label{sec:discussion}
\begin{figure}[ht!]
  \centering
  \includegraphics[width=0.85\linewidth]{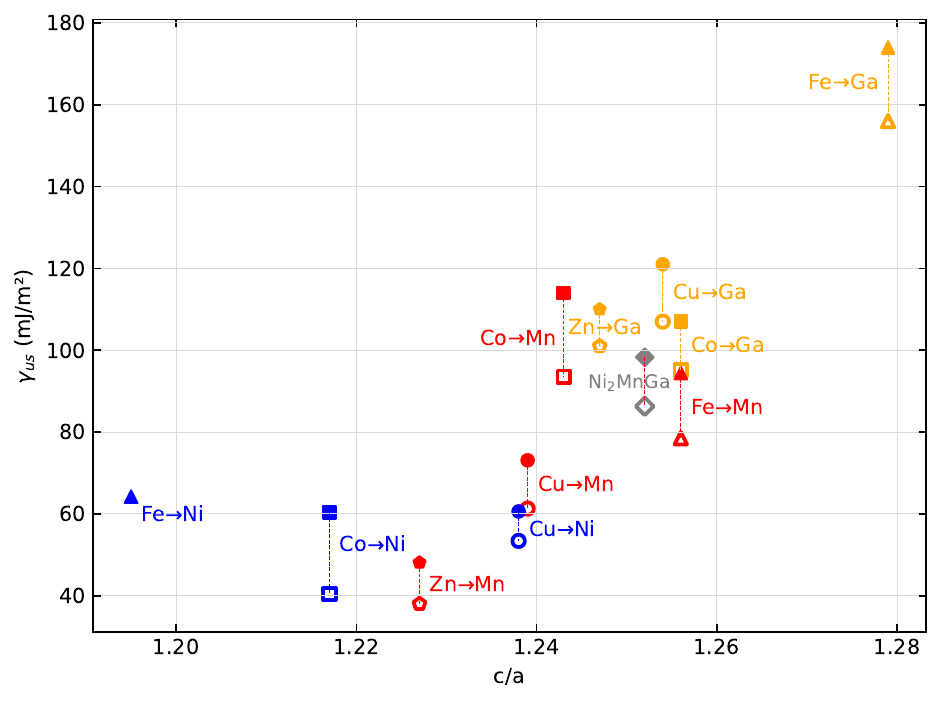}
  \caption{First GPFE maximum, $\gamma _{us}$, for the doped systems and stoichiometric Ni$_{2}$MnGa as a function of the tetragonal ratio $c/a$. Filled and open symbols denote non-optimized and optimized values, respectively.}
  \label{fig:firstMax}
\end{figure}

If we omit Fe$\rightarrow$Ni substitution, which results in suppression of the martensitic transformation~\cite{Kopecky2021-jk} and produces unusual behavior of the GPFE curve, the effects of individual dopants can be divided into three groups: (i) Cu$\rightarrow$Mn, Cu$\rightarrow$Ni, Co$\rightarrow$Ni, and Zn$\rightarrow$Mn substitutions, which reduce the first barrier $\gamma _{us}$ and support twin formation; (ii) Cu$\rightarrow$Ga, Co$\rightarrow$Mn, Co$\rightarrow$Ga, Fe$\rightarrow$Ga, and Zn$\rightarrow$Ga substitutions, which increase barriers along the GPFE curve and hinder twin formation or propagation; and (iii) Fe$\rightarrow$Mn substitution, which has only a minor effect on the barriers. If we plot $\gamma _{us}$ as a function of $c/a$ for studied alloys (Fig.~\ref{fig:firstMax}), we indeed see that Cu$\rightarrow$Mn, Cu$\rightarrow$Ni, Co$\rightarrow$Ni, and Zn$\rightarrow$Mn substitutions result in a significant decrease of $c/a$ compared to the stoichiometric value 1.252. This is consistent with the experimental observation that only NM martensite with reduced $c/a$ exhibits low twinning stress~\cite{Soroka2018-ci}. Cu$\rightarrow$Ga and Co$\rightarrow$Ga result in only a small change in $c/a$, whereas Fe$\rightarrow$Ga even increases it. The Co$\rightarrow$Mn and Zn$\rightarrow$Ga substitutions should be noted as exceptions, because they increase the first barrier for twin nucleation while still decreasing $c/a$, although the decrease is very small. Numerical values of $\gamma _{us}$ and $c/a$ are listed in Table~\ref{tab:gpfe_peaks}.

Concerning transformation temperatures, there appears to be no clear correlation between a dopant’s effect on the GPFE curve and its effect on the experimentally determined trends in $T_c$ and $T_m$ summarized in Table~\ref{tab:configurations} (i.e., on the stability of ferromagnetic ordering and the martensitic phase). For example, Cu$\rightarrow$Mn and Cu$\rightarrow$Ni both decrease the initial barrier, but they shift the transformation temperatures in opposite directions: Cu$\rightarrow$Mn increases $T_{m}$ and decreases $T_{c}$, whereas Cu$\rightarrow$Ni decreases $T_{m}$ and increases $T_{c}$. However, with the exception of Cu$\rightarrow$Mn, dopants that stabilize the martensitic phase tend to increase the barrier and hinder twin nucleation.

A useful energetic interpretation of Fig.~\ref{fig:firstMax} is that $\gamma_{us}$ represents the excess energy required for twin nucleation (formation of the intrinsic stacking fault) along the prescribed shear pathway. For the examined shear system, this excess energy contains a substantial mechanical contribution governed by the lattice geometry and shear stiffness. Therefore, substitutions that reduce the tetragonal distortion require a smaller twinning shear and are generally accompanied by a reduction of the corresponding shear modulus, thereby lowering the mechanical energy accumulated along the shear pathway. This is consistent with the observed decrease of $\gamma_{us}$ for substitutions that significantly lower $c/a$ and with the general requirement of low twinning stress for MIR. However, the GPFE barrier is not determined by this mechanical contribution alone, since the local chemical and magnetic environment at the shearing plane can also modify the energetics of stacking-fault formation.

\begin{figure}[ht!]
    \centering
    \includegraphics[width=0.75\linewidth]{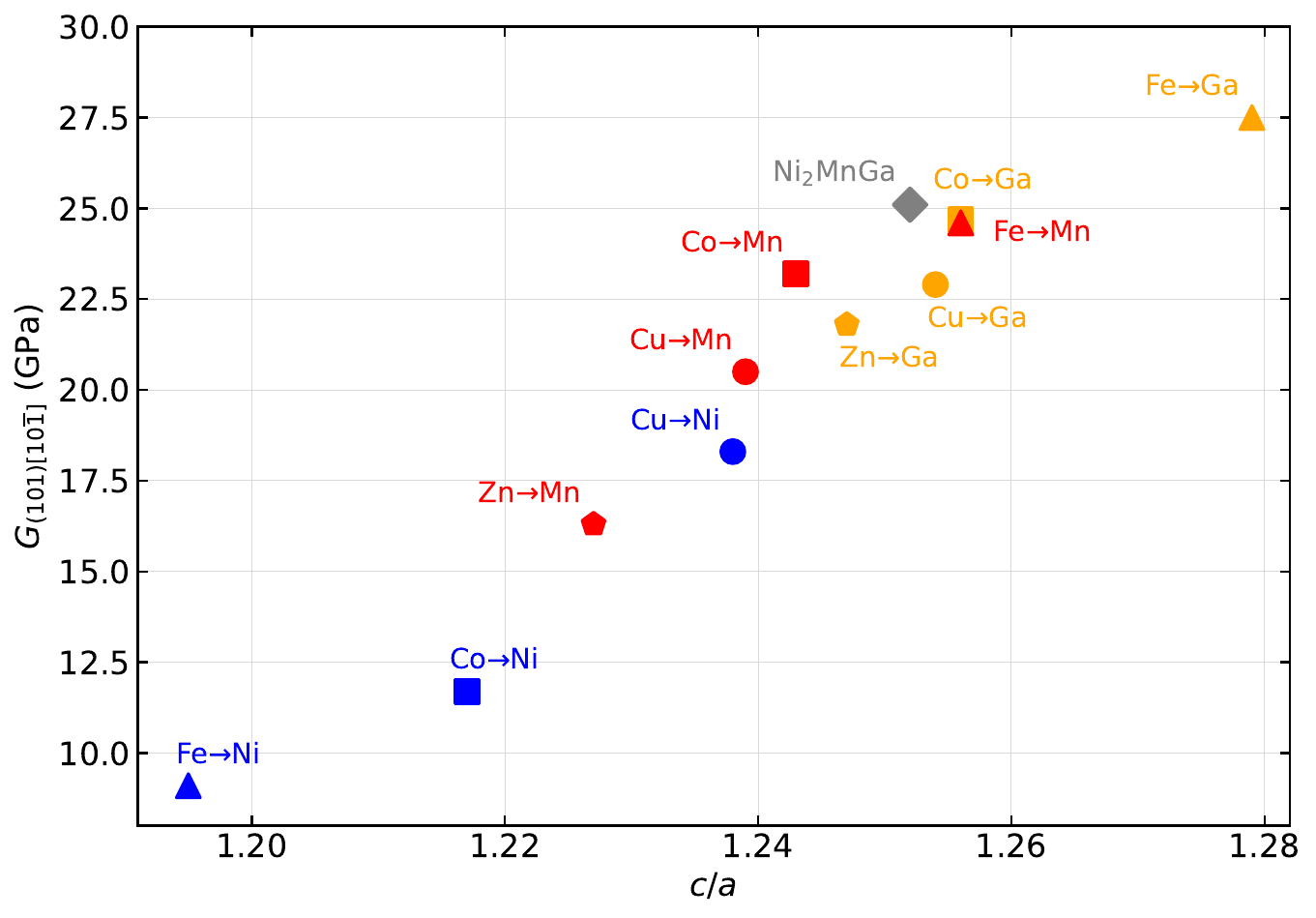}
    \caption{Shear modulus $G_{(101)[10\bar{1}]}$ of the considered systems as a function of the tetragonal ratio $c/a$.}
    \label{fig:shear_modulus_ca}
\end{figure}

This interpretation also helps to clarify the substitutions that deviate from the main trend. For Co$\rightarrow$Mn and Zn$\rightarrow$Ga, the decrease in $c/a$ is small, whereas $\gamma_{\mathrm{us}}$ increases. To determine whether these deviations could be attributed to differences in lattice stiffness, we calculated the shear modulus $G_{(101)[10\bar{1}]}$ for all investigated configurations. The calculated values are summarized in Table~\ref{tab:gpfe_barriers} and plotted as a function of $c/a$ in Fig.~\ref{fig:shear_modulus_ca}. The shear modulus exhibits a nearly linear dependence on $c/a$, including the results for Co$\rightarrow$Mn and Zn$\rightarrow$Ga. Cu$\rightarrow$Ga shows a slightly lower modulus than expected from its $c/a$, which remains close to that of stoichiometric Ni$_2$MnGa; however, this deviation is small. These results indicate that lattice geometry is the principal factor controlling the shear stiffness. Consequently, the anomalous increase of $\gamma_{us}$ for Co$\rightarrow$Mn and Zn$\rightarrow$Ga cannot be explained by the shear modulus and is instead attributed to local chemical and magnetic effects associated with the particular atomic configuration at the shearing plane.

An additional trend emerging from Figs.~\ref{fig:firstMax} and \ref{fig:shear_modulus_ca} is the systematic grouping according to the substituted sublattice. Substitutions on the Ni sublattice, shown in blue, are located predominantly at lower $c/a$, lower shear modulus, and lower $\gamma_{us}$, indicating generally more favorable conditions for twin nucleation. In contrast, substitutions on the Ga sublattice, shown in yellow, tend to occupy the opposite region, with higher $c/a$, higher shear stiffness, and increased $\gamma_{us}$, corresponding to a less favorable twinning response. Substitutions on the Mn sublattice, shown in red, are generally located closer to stoichiometric Ni$_{2}$MnGa but tend overall toward lower $c/a$, lower shear stiffness, and reduced $\gamma_{us}$, indicating a weaker tendency to facilitate twinning. Co$\rightarrow$Mn represents the main exception to this trend, since its $\gamma_{us}$ increases despite the modest reductions in $c/a$ and shear modulus.

Figures \ref{fig:main} and \ref{fig:Zn} compare the GPFE curves with the stoichiometric Ni$_{2}$MnGa NM martensite reference. However, in the stoichiometric alloy the austenite transforms to modulated 10M martensite, and NM martensite has not been reported for this composition. Therefore, our results should also be compared with the recently published GPFE curve for off-stoichiometric NM martensite with composition Ni$_{50}$Mn$_{28.125}$Ga$_{21.875}$~\cite{Heczko2026-vx} which is displayed in Fig.~\ref{fig:GPFE-comaprison} (b). The Cu$\rightarrow$Ga, Co$\rightarrow$Mn, Co$\rightarrow$Ga, and Fe$\rightarrow$Ga substitutions result in an even larger increase of GPFE barriers than that caused by increased Mn content. Therefore, for these substitutions, a stronger negative effect on twin nucleation and propagation can be expected in NM martensite for the same dopant concentration as for excess Mn. Among the considered Ga-site substitutions, only Zn$\rightarrow$Ga exhibits a weaker effect. 

\begin{figure}[t]
    \centering
    \includegraphics[width=\linewidth]{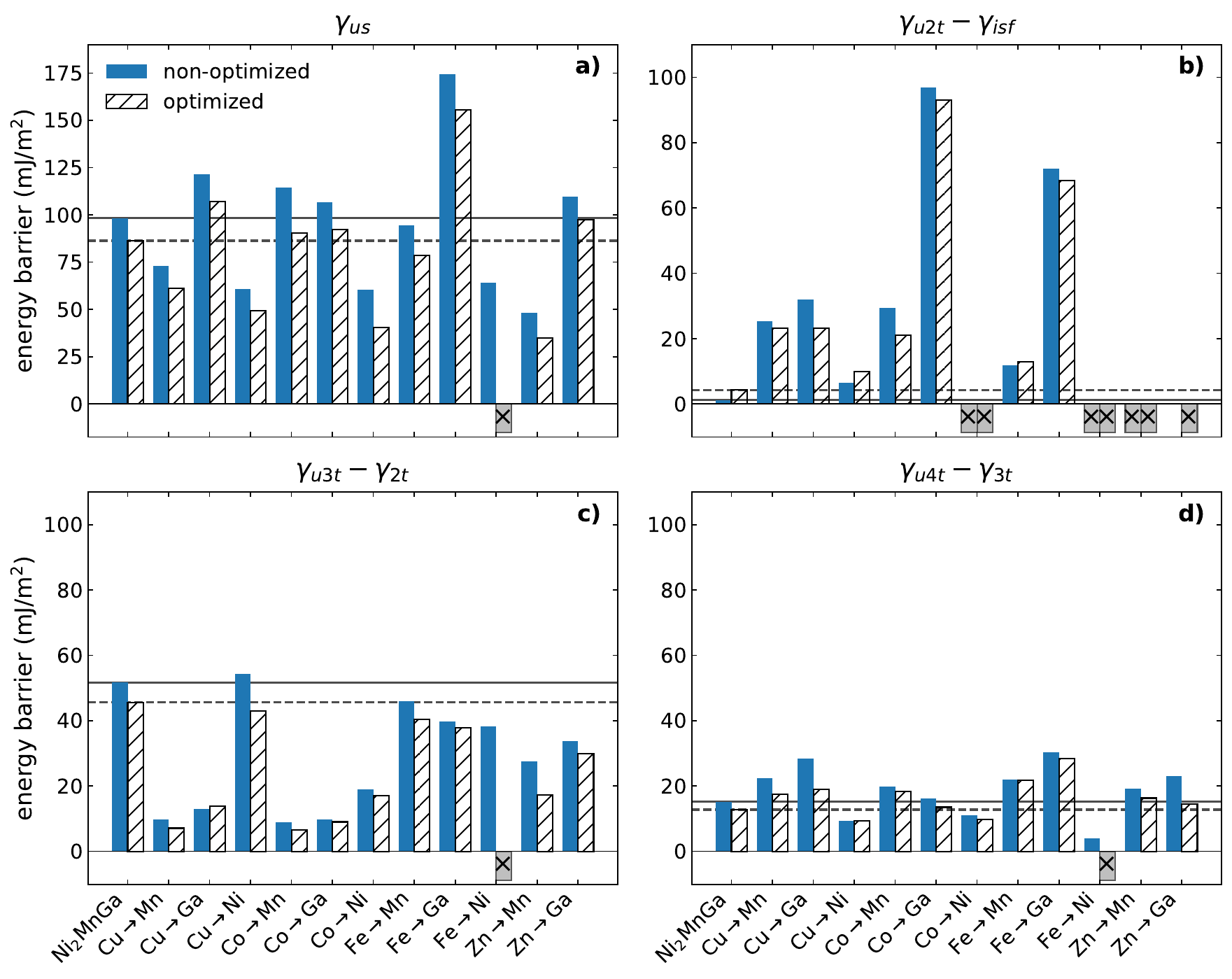}
    \caption{Comparison of the energy barriers extracted from the generalized planar fault
energy (GPFE) curves before and after G-SSNEB optimization. The four successive barriers correspond to (a) $\gamma_{us}$, (b) $\gamma_{u2t}-\gamma_{isf}$, (c) $\gamma_{u3t}-\gamma_{2t}$, and (d) $\gamma_{u4t}-\gamma_{3t}$. A cross ($\times$) indicates that the corresponding barrier could not be determined because the respective maximum is absent from the GPFE curve. Horizontal lines indicate the corresponding barrier values for stoichiometric Ni$_2$MnGa, with solid and dashed lines representing
the non-optimized and optimized values, respectively. Note the different y-axis scale used in panel (a) compared with panels (b)--(d).}
    \label{fig:barriers}
\end{figure}

As the height of the first barrier $\gamma _{us}$ is associated with twin nucleation and formation of the intrinsic stacking fault, the subsequent barriers indicate the difficulty of further twin propagation. Thus, an “ideal” GPFE curve corresponding to low twinning stress should exhibit a small first maximum followed by a smooth profile with shallow minima and no pronounced maxima. To facilitate a quantitative comparison of these features among the considered substitutions, the heights of the successive barriers extracted from the GPFE curves are summarized in Table~\ref{tab:gpfe_barriers} and Fig.~\ref{fig:barriers}. The "ideal" behavior is observed for Cu$\rightarrow$Mn substitution and, to some extent, also for Co$\rightarrow$Ni substitution. Notably, these substitutions correspond to the alloy with nominal composition Ni$_{46}$Co$_{4}$Mn$_{24}$Ga$_{22}$Cu$_{4}$, which exhibits the largest reported magnetically induced strain of 12\% and a twinning stress low enough to allow MIR~\cite{Sozinov2013-si}. Although the nominal composition suggests rather Cu$\rightarrow$Ga substitution, which in our results increases the barrier for twin formation, the strong evaporation of Mn during alloy casting~\cite{Zeleny2020-ll} may lead to a higher effective Cu content on the Mn sublattice. Thus, proper balancing of Cu substitution between the Ga sublattice (increasing T$_{m}$) and the Mn sublattice (facilitating twin propagation) appears to be key for the design of this alloy. However, it should be noted that the effect of combined doping can differ from the trends observed for individual substitutions, even though their effects on $T_{m}$, $T_{c}$, and $c/a$ are reported to be rather independent~\cite{Zeleny2014}.

Despite these limitations, the available experimental data provide qualitative support for the predicted site-dependent trends. Sozinov et al. reported a twinning stress of approximately 15–20 MPa for the Mn-excess NM single crystal Ni$_{52.1}$Mn$_{27.3}$Ga$_{20.6}$ with $c/a\approx1.21$~\cite{Sozinov2001-lf}, whereas the Co–Cu-doped NM single crystals studied by Soroka et al. exhibited substantially lower room-temperature values of approximately 0.8–1.4 MPa~\cite{Soroka2018-ci}. The low twinning stresses of the Co–Cu-doped single crystals are consistent with the reduced tetragonality and shear stiffness of these alloys and with the predicted barrier-lowering effect of Co$\rightarrow$Ni substitution. Within the Soroka et al. series, replacing 0.5 at.\% Ga by Cu increased the twinning stress from 0.77 to 1.37 MPa, while replacing 1 at.\% Mn by Co increased it from approximately 1.27 to 1.41 MPa. These trends agree qualitatively with the increased GPFE barriers calculated for Cu$\rightarrow$Ga and Co$\rightarrow$Mn substitutions. The relatively modest changes within this series should be viewed in the context of an already very low baseline twinning stress, which is largely attributed to the beneficial effect of Co addition. Brzoza-Kos et al.~\cite{Brzoza-Kos2022-wm} likewise reported high twinning stresses in Cu-rich polycrystalline NM martensite, further supporting the unfavorable effect of Cu substitution on the Ga sublattice, although the absolute values cannot be compared directly with single-crystal measurements because of grain-boundary constraints.

\begin{table*}[t]
\centering
\caption{Calculated GPFE barrier heights together with the shear modulus
$G_{(101)[10\overline{1}]}$. The first barrier, $\gamma_{\mathrm{us}}$,
corresponding to twin nucleation is reported in Table~II to avoid duplication.
Values obtained after optimization are given in parentheses. N/A indicates that
the corresponding barrier could not be determined because the respective maximum
is absent from the GPFE curve.}
\label{tab:gpfe_barriers}

\renewcommand{\arraystretch}{1.15}
\setlength{\tabcolsep}{5pt}

\begin{tabular}{l c c c c}
\toprule
substitution &
$G_{(101)[10\overline{1}]}$ &
$\gamma_{\mathrm{u2t}}-\gamma_{\mathrm{isf}}$ &
$\gamma_{\mathrm{u3t}}-\gamma_{\mathrm{2t}}$ &
$\gamma_{\mathrm{u4t}}-\gamma_{\mathrm{3t}}$ \\
&
(GPa) &
\multicolumn{3}{c}{(mJ\,m$^{-2}$)} \\
\midrule
Ni$_2$MnGa        & 25.1 & 1.2 (4.3)    & 51.7 (45.6)    & 15.2 (12.8) \\
Cu$\rightarrow$Mn & 20.5 & 25.2 (23.1)  & 9.8 (7.1)    & 22.4 (17.6) \\
Cu$\rightarrow$Ga & 22.9 & 31.9 (23.3)  & 13.0 (13.9)    & 28.4 (19.0) \\
Cu$\rightarrow$Ni & 18.3 & 6.4 (10.0)   & 54.3 (43.0)    & 9.3 (9.4) \\
Co$\rightarrow$Mn & 23.2 & 29.5 (21.0)  & 8.8 (6.5)    & 19.8 (18.4) \\
Co$\rightarrow$Ga & 24.7 & 96.8 (93.1)  & 9.8 (9.0)    & 16.2 (13.5) \\
Co$\rightarrow$Ni & 11.7 & N/A (N/A)    & 19.0 (17.1)   & 11.1 (9.73) \\
Fe$\rightarrow$Mn & 24.6 & 11.9 (12.9)  & 45.9 (40.4)    & 21.9 (21.8) \\
Fe$\rightarrow$Ga & 27.5 & 71.9 (68.3)  & 39.7 (37.7)    & 30.3 (28.5) \\
Fe$\rightarrow$Ni & 9.1  & N/A          & 38.4         & 4.0  \\
Zn$\rightarrow$Mn & 16.3 & N/A (N/A)          & 27.5 (17.3)    & 19.2 (16.3) \\
Zn$\rightarrow$Ga & 21.8 & 0.2 (N/A)       & 33.7 (29.9)    & 23.1 (14.5) \\
\bottomrule
\end{tabular}

\end{table*}

In our previous work~\cite{Heczko2024-of}, we associated the second minimum $\gamma _{2t}$ with the ability of the structure to form lattice modulation. Therefore, doping that deepens the second minimum is expected to further stabilize the modulated structure. Such behavior is observed for Fe$\rightarrow$Mn substitution, which indeed exhibits an extended concentration range of stable modulated martensites, and also for Zn$\rightarrow$Mn substitution, for which corresponding experimental evidence is still lacking. Experimental preparation of Zn-doped alloys with partial Zn occupancy of the Mn sublattice would therefore provide a valuable test of the predicted relationship between the depth of the second GPFE minimum $\gamma _{2t}$ and the stability range of modulated martensites. A similar but much stronger effect is also seen for Cu$\rightarrow$Ni substitution, where even the structure containing the intrinsic stacking fault becomes more stable than the initial defect-free structure; however, it is accompanied by a decrease of $T_{m}$ and an overall destabilization of martensite~\cite{Kanomata2005-hj}. On the other hand, the energy of the second minimum $\gamma_{2t}$ increases significantly for Co$\rightarrow$Mn and Co$\rightarrow$Ga substitutions, which corresponds to strong destabilization of modulated phases and stabilization of NM martensite. Recent experimental results confirm such behavior, although the experimentally prepared alloy exhibits much more complex site occupancy than our simplified model~\cite{RIOLOPEZ2026186643}.

Zn$\rightarrow$Ga substitution has been proposed as an alternative to Cu$\rightarrow$Ga substitution, as it also increases $T_{m}$~\cite{Barton2014-ox}  but has a less negative effect on $T_{c}$ and magnetocrystalline anisotropy~\cite{Janovec2020-ko}. A comparison of the GPFE curves for these two cases in Figures 4 and 5 clearly reveals that the first maximum—and especially the second maximum—are significantly lower for Zn$\rightarrow$Ga substitution than for Cu$\rightarrow$Ga substitution. These results do not necessarily imply that Cu should be completely replaced by Zn in a simple quaternary alloy. Instead, they suggest that partial replacement of Cu by Zn could be considered in the established Cu–Co-doped Ni–Mn–Ga alloy. In such a multi-doped system, Zn should preferentially occupy the Ga sublattice, where it could retain the beneficial increase in $T_m$ while reducing the detrimental effects of Cu$\rightarrow$Ga substitution on $T_c$, magnetocrystalline anisotropy, and the GPFE barriers. Co would retain its role in reducing tetragonality and twinning stress and in partly compensating for the reduction of $T_c$. Thus, Zn$\rightarrow$Ga substitution may provide a useful component of a balanced multi-element alloying strategy, although the actual site occupancy and the combined effects of Cu, Co, and Zn must be verified experimentally. 

The experimental verification of these predictions may nevertheless be challenging because the high volatility of Zn complicates conventional melting-based preparation of Zn-containing Ni–-Mn–-Ga-based alloys with increased Zn content~\cite{Barton2014-ox}. This limitation might potentially be overcome by alternative fabrication routes that minimize or avoid melting. Possible approaches include mechanical alloying of powder components followed by spark plasma sintering and annealing, solid-state deposition followed by controlled diffusion heat treatment, or cold-spray deposition using high-velocity Zn-containing particles. In all cases, subsequent heat treatment would likely be required to promote homogenization and chemical ordering. These approaches should therefore be regarded as possible directions for future experimental studies rather than established preparation routes for Zn-containing Ni–-Mn–-Ga-based alloys.

The interpretation of Zn$\rightarrow$Mn substitution is more limited. Although the presence of Zn on the Mn sublattice decreases all maxima along the GPFE curve and may therefore facilitate twin nucleation and propagation, the profile is less smooth than in the case of Cu$\rightarrow$Mn substitution. Moreover, available first-principles predictions indicate that Zn$\rightarrow$Mn substitution decreases both $T_m$ and $T_c$. These changes may offset, or even outweigh, the favorable reduction of the GPFE barriers. Therefore, Zn$\rightarrow$Mn should not be regarded as an optimal stand-alone alloying strategy. Nevertheless, partial Zn occupancy of the Mn sublattice may still have a beneficial effect in a complex Cu–Co–Zn-doped alloy, analogously to the proposed role of Cu$\rightarrow$Mn substitution in the Cu–Co-doped composition. The net effect of such combined doping on the transformation temperatures, magnetic properties, and twinning behavior cannot be inferred from the present single-substitution calculations and must be assessed experimentally.

\section{Conclusions}

In the present work, we use first-principles calculations based on density functional theory to determine the energy profiles along GPFE curves for the $(101)[10\bar{1}]$ shear system in Ni–Mn–Ga NM martensite doped with Cu, Co, Fe, and Zn on different sublattices. By comparing the calculated GPFE profiles with the profile of stoichiometric Ni$_{2}$MnGa, we estimate the qualitative effect of dopants on twinning stress and the MIR effect in NM martensite. This interpretation is based on the fact that the first barrier on the curve $\gamma_{us}$ is associated with formation of the intrinsic stacking fault, i.e., twin nucleation, whereas subsequent barriers correspond to obstacles for twin-boundary motion, i.e., twin growth. We find that Cu$\rightarrow$Mn, Cu$\rightarrow$Ni, Co$\rightarrow$Ni, and Zn$\rightarrow$Mn substitutions result in a decrease of the first barrier and support twin nucleation. These substitutions also significantly decrease the tetragonal ratio $c/a$ of the NM structure, which supports the relation between low twinning stress and decreased $c/a$. In contrast, Cu$\rightarrow$Ga, Co$\rightarrow$Mn, Co$\rightarrow$Ga, Fe$\rightarrow$Ga, and Zn$\rightarrow$Ga substitutions increase the barriers along the path and make twin formation more difficult. However, the same substitutions are used to increase T$_{m}$ because they stabilize the martensitic phase. The effect of Fe→Mn substitution appears to be negligible, as there is no significant difference between the GPFE curves for the doped and stoichiometric alloys. The effect of Fe$\rightarrow$Ni substitution is difficult to assess because the calculated energy profile exhibits unusual features consistent with unstable twin configurations. The calculated trends also reveal a clear sublattice dependence: substitutions on the Ni sublattice tend to reduce $c/a$, shear stiffness, and $\gamma_{us}$, whereas substitutions on the Ga sublattice show the opposite tendency; substitutions on the Mn sublattice generally lie closer to stoichiometric Ni$_{2}$MnGa, with a weaker overall tendency toward reduced values.

Because the modulation of 10M and 14M martensite can be described as nanotwinning of the NM lattice, we also use the GPFE curve to assess modulation stability. In particular, deepening of the second minimum on the GPFE curve (i.e., for the two-layer nanotwin) caused by doping can be interpreted as stabilization of lattice modulation, since the two-layer nanotwin is a common structural motif for both 10M and 14M martensite. We find the second minimum to be energetically lower than the defect-free structure for Fe$\rightarrow$Mn, Cu→Ni, and Zn$\rightarrow$Mn substitutions, which indicates a potentially increased concentration range in which modulated martensite could be stable. In the other considered substitutions, the energy of the second minimum increases compared to the initial defect-free structure, which indicates a tendency to destabilize modulation. Overall, our findings offer guidance for experimental efforts aimed at selecting promising dopants and compositions to optimize functional properties of magnetic shape memory alloys.


\section*{Data availability statement}
The data supporting the findings of this study are publicly available from Zenodo at https://doi.org/10.5281/zenodo.21998433.

\section*{Acknowledgements}

The authors acknowledge the funding support from the Ministry of Education, Youth and Sports of the Czech Republic (Project No. LUC25051 and Ferroic Multifunctionalities project (FerrMion) No. CZ.02.01.01/00/22\_008/0004591, co-funded by the European Union), from the Czech Science Foundation (Project No. 24-10334S) and from the National Science Centre of Poland (Project No. 2021/42/E/ST5/00367). Computational resources were provided under the Projects e-INFRA CZ (ID:90254) at the IT4Innovations National Supercomputing Center - www.it4i.cz.

\bibliographystyle{ieeetr}
\bibliography{Ni2MnGa.bib}

\end{document}